\documentclass[proof]{pasj00}
\usepackage{color}

\begin{document}
\SetRunningHead{A. Asai et al.}{Red Asymmetry of Flare Ribbons}
\Received{2009/07/01}
\Accepted{2011/09/20}

\title{A Study on Red Asymmetry of H$\alpha$ Flare Ribbons Using
Narrowband Filtergram in the 2001 April 10 Solar Flare}

\author{
Ayumi \textsc{Asai}\altaffilmark{1}, 
Kiyoshi \textsc{Ichimoto}\altaffilmark{2}
Reizaburo \textsc{Kitai}\altaffilmark{2}, 
Hiroki \textsc{Kurokawa}\altaffilmark{2}, 
 and
Kazunari \textsc{Shibata}\altaffilmark{2}}

\altaffiltext{1}{Unit of Synergetic Studies for Space, Kyoto University,
Yamashina, Kyoto, 607-8471, Japan}
\email{asai@kwasan.kyoto-u.ac.jp}
\altaffiltext{2}{Kwasan and Hida Observatory, Kyoto University,
Yamashina, Kyoto, 607-8471, Japan}


\maketitle

\begin{abstract}
We report a detailed examination of the ``red asymmetry'' of H$\alpha$
emission line seen during the 2001 April 10 solar flare by using a
narrowband filtergram.
We investigated the temporal evolution and the spatial distribution of
the red asymmetry by using the H$\alpha$ data taken with the 60cm
Domeless Solar Telescope at Hida Observatory, Kyoto University.
We confirmed that the red asymmetry clearly appeared all over the flare
ribbons, and the strong red asymmetry is located on the outer narrow
edges of the flare ribbons, with the width of about $1.^{\prime\prime}5$
-- $3.^{\prime\prime}0$ (1000 -- 2000~km), where the strong energy
releases occur.
Moreover, we found that the red asymmetry, which also gives a measure of
the Doppler shift of the H$\alpha$ emission line concentrates on a
certain value, not depending on the intensity of the H$\alpha$ kernels.
This implies not only that the temporal evolutions of the red asymmetry
and those of the intensity are not in synchronous in each flare kernel,
but also that the peak asymmetry (or velocity of the chromospheric
condensation) of individual kernel is not a strong function of their
peak intensity.
\end{abstract}

\section{Introduction}
Solar flares have been extensively observed in H$\alpha$ line, since
they show spectacular phenomena in this wavelength, such as
filament/prominence eruptions, flare ribbons, post flare loops (or
coronal rains), and so on (e.g. \cite{Tan88}).
H$\alpha$ kernels, very bright and compact H$\alpha$ emission sources
seen in flare ribbons, are also one of the prominent features that
appear during a flare.
Such flare ribbons and kernels have been thought to be caused by the
energy deposition from the corona into the chromosphere.

To explain these various phenomena successfully, magnetic reconnection
models have been widely discussed for last several decades.
Especially, the so-called CSHKP magnetic reconnection model
\citep{Car64,Stur66,Hira74,Kopp76} is often accepted as the ``standard''
model.
In the scenario, the connectivity of magnetic field lines is
topologically changed (i.e. re-connected) in the corona, and then, vast
magnetic energy stored in the corona is released.
The released energy is converted to various kinds of energy; kinetic
energy spent for ejections of plasma, thermal energy to generate hot
plasma of about 10~MK or more, nonthermal energy to accelerate energetic
particles, and so on.
The thermal conduction and/or the nonthermal particles rapidly travel
along newly connected magnetic field lines.
Then, they reach the chromosphere, followed by the bombardment to the
chromospheric plasma.
H$\alpha$ flare ribbons and kernels are caused by such bombardment.
The CSHKP model further suggests that magnetic field lines successively
reconnect in the corona.
During the successive magnetic reconnection, the reconnection points
(X-points) move upward.
As a result, newly reconnected field lines have their footpoints farther
out from the neutral line than the footpoints of the field lines that
have already reconnected.
Therefore, the separation of the flare ribbons is not a real motion, but
an apparent one (see also the reviews of the CSHKP model by
\cite{Stur92} and \cite{Sve92}).

The energy deposition into the chromosphere also causes strong
distortions of H$\alpha$ line (line width and line shift) besides the
emission.
The enhancements of the intensity at red wings are particular, and have
been reported as ``red asymmetry'' of the lines \citep{Sve76}, while
``blue asymmetry'' is also reported to be sometimes seen at very early
phase of a flare.
Comparing between the red and the blue wing images, it has known that
such red asymmetry is seen overall the flare ribbons
\citep{Jan70,Sve76,Tang83}.
\citet{Hana03} examined the H$\alpha$ impact polarization (liner
polarization) due to collisions of high-energy particles at flare
kernels, and reported that red asymmetry also appears at such kernels
during the rise phase of kernel brightenings.
Observations on the red asymmetry with filtergraphs, however, have not
been progressed so much since then, and the further characteristics on
the spatial distribution of the red asymmetry remains to be revealed.

\citet{Ichi84} reported that the red-shift corresponds to the downward
motion with speed of several tens of km~s$^{-1}$, based on their
excellent spectroscopic observations.
\citet{Fal97} further reported that strong red asymmetry, which
corresponds to the downward velocity of the order of tens of
km~s$^{-1}$, is observed at the outer edge of the flare ribbons.
The width of the edges is about 2 -- 3$^{\prime\prime}$ ($\sim$ 1500 --
2000 km).
This is reasonable in the context of magnetic reconnection theories
suggesting that strong energy depositions occur at the outer edges of
the flare ribbons connected by the newly reconnected loops.
\citet{Ichi84} also quantitatively examined the red asymmetry.
The strong enhancement of the chromospheric plasma due to the
precipitation of nonthermal plasma and/or thermal conduction causes not
only red asymmetry of the chromospheric lines but explosive upflow of
the heated plasma along the flare loops.
The latter is called ``chromospheric evaporation'', and has been
observed as blue-shift of coronal emission lines \citep{Hira74,Anto84}.
\citet{Ichi84} found a momentum balance between the downward motion of
the chromospheric condensation and the chromospheric evaporation, and
confirmed that the red asymmetry is a counteraction of the chromospheric
evaporation.

Red asymmetry, in this way, tells us the quantitative information of the
precipitation (nonthermal particles / thermal conduction).
Understanding the features of the chromospheric plasma at the footpoints
of flare loops is strongly related with understanding energy
release/deposit mechanisms and is crucially important for solar flare
studies.
It has been, however, a challenge theoretically and numerically to
explain exactly those observational facts (e.g. \cite{Fish85a,Fish85b}),
since the distortion of chromospheric lines is too complicated.
Observations on the red asymmetry, have not been progressed so much
since \citet{Ichi84}.
Especially, spatial distribution of the red asymmetry using narrowband
filtergram, remains to be revealed, although \citet{Tang83} reported
that red asymmetry features appear all over flare ribbons and continue
from onset until late in the decay phase by using narrowband filtergram
of H$\alpha$ line.
In this paper we examined the spatial distribution of the red asymmetry
at H$\alpha$ kernels and the temporal evolution, by using H$\alpha$ wing
images.
In \S~2 we summarize the observational data and the features of the
flare.
In \S~3 we show the features of the red asymmetry.
In \S~4 a summary and discussions are given.

\section{Observations}
We observed a typical two-ribbon flare (X2.3 on the GOES scale) that 
occurred in the NOAA Active Region 9415 (S22$^{\circ}$, W01$^{\circ}$)
at 05:10 UT, 2001 April 10 with the 60~cm Domeless Solar Telescope
(DST\footnote{http://www.kwasan.kyoto-u.ac.jp/general/facilities/dst/index\_en.html})
at Hida Observatory, Kyoto University.
The H$\alpha$ monochromatic images were obtained with the Zeiss Lyot
filter of the 0.25~{\AA} passband and KODAK 4.2{\it i} Megaplus CCD
camera with the size of 2024 $\times$ 2040 pixels sequentially in eight 
wavelengths: H$\alpha$ $-1.5$, $-0.8$, $-0.4$, $\pm0.0$, $+0.4$, $+0.8$,
$+1.5$, and $+5.0$~{\AA}.
The successive wavelength change and recording were controlled with a
personal computer, and the time cadence for each wavelength was about
30~s.
The pixel size of the data is $0.^{\prime\prime}28$ (the highest
resolution of the system is $0.^{\prime\prime}14$, but binning was done
in the current case).
However, the spatial resolution is about $2^{\prime\prime}$ or worse
according to the seeing.

In Figure~1 we present snap shots of the H$\alpha$ images in the (a)
$-1.5$, (b) $-0.8$, (c) $-0.4$, (d) $\pm 0$ (line center), (e) $+1.5$,
(f) $+0.8$, (g) $+0.4$, and (h) $+5.0$ {\AA} wings, respectively.
During the observation, the emissions at H$\alpha$ kernels were so
strong that some fine structures inside the flare ribbons were missed
suffering from the saturation in the core.
In this study, therefore, we mainly used the H$\alpha$ $-1.5$ and
$+1.5$~{\AA} images only, in which the flare kernels are clearly seen
rarely suffering from the saturation.
In Figure~2 we show the temporal development of the flare in the
H$\alpha$ center ({\it top}), in the $+0.8$ {\AA} wing ({\it middle}),
and in the $-0.8$ {\AA} wing ({\it bottom}).
The filament eruption is clearly seen in the blue wing images before the
two ribbon flare starts.
As the flare progresses the two ribbon structure evolves and separates
to each other.
In the later phase, downflows along the post flare loops (coronal rains)
are also seen in red wing images.

As Figure~1 shows and \citet{Asai04} reported, the two flare ribbons lie
roughly in the north-south direction, and the east (left) ribbon is of
positive magnetic polarity, and the west (right) one is of negative
polarity.
As the flare progresses, the flare ribbons separate to each other, in
the horizontal (that is, in the east-west) direction.
We already reported the characteristics of the flare in H$\alpha$,
which was observed with the {\it Sartorius} Telescope at Kwasan
Observatory, Kyoto University, in some previous papers
\citep{Asai02,Asai03,Asai04}.
We also quantitatively discussed the relation between the flare ribbon
evolution and the energy release rate in these papers.
\citet{Xie09} also studied the ribbon separation of this flare.
In this event the oscillation of an H$\alpha$ filament driven by an EIT
wave is also detected in a distant location \citep{Oka04}.

\begin{figure}
  \begin{center}
    \FigureFile(90mm,180mm){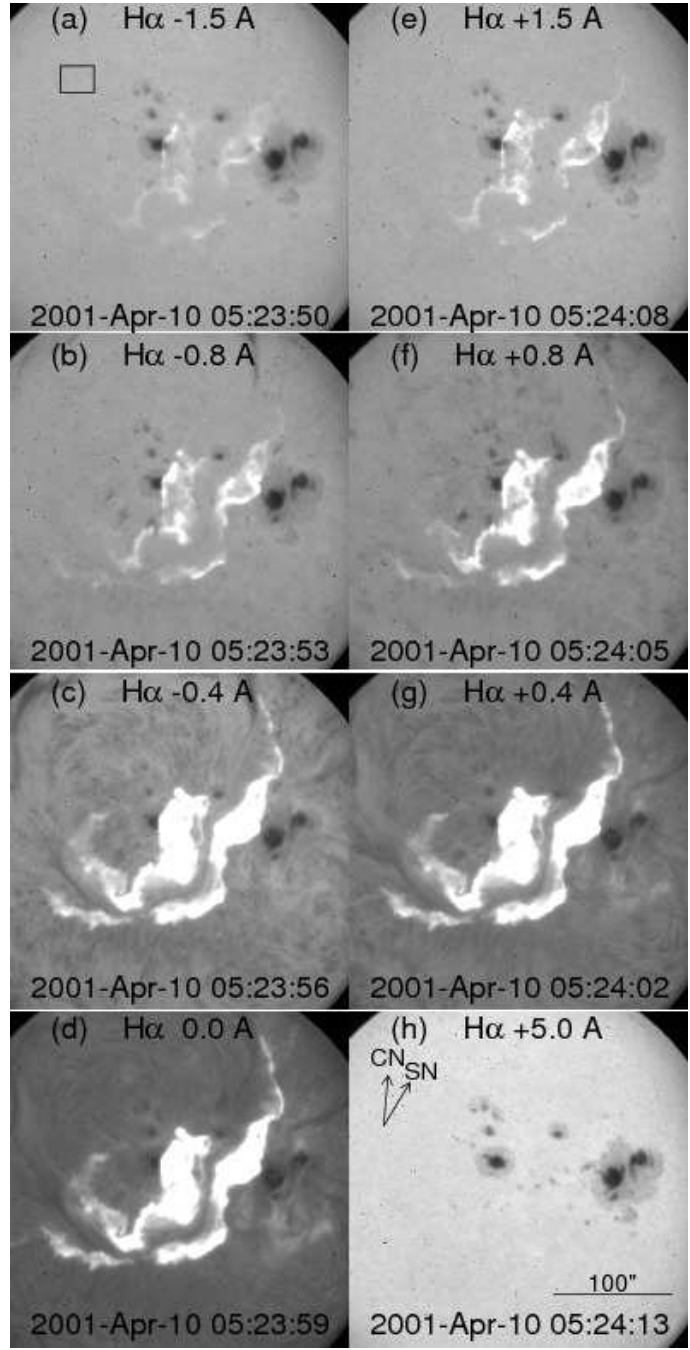}
  \end{center}
  \caption{
H$\alpha$ images of the 2001 April 10 flare in the (a) $-1.5$, (b)
$-0.8$, (c) $-0.4$, (d) $\pm 0$ (line center), (e) $+1.5$, (f) $+0.8$,
(g) $+0.4$, and (h) $+5.0$ {\AA} wings of H$\alpha$ line.
Celestial north is up as shown with the arrow denoted ``CN'' in the
panel (f), and west is to the right.
The direction of the solar north is also shown with the arrow denoted
``SN'' in the panel (h).
}\label{fig:dst}
\end{figure}

\begin{figure}
  \begin{center}
    \FigureFile(162.5mm,120mm){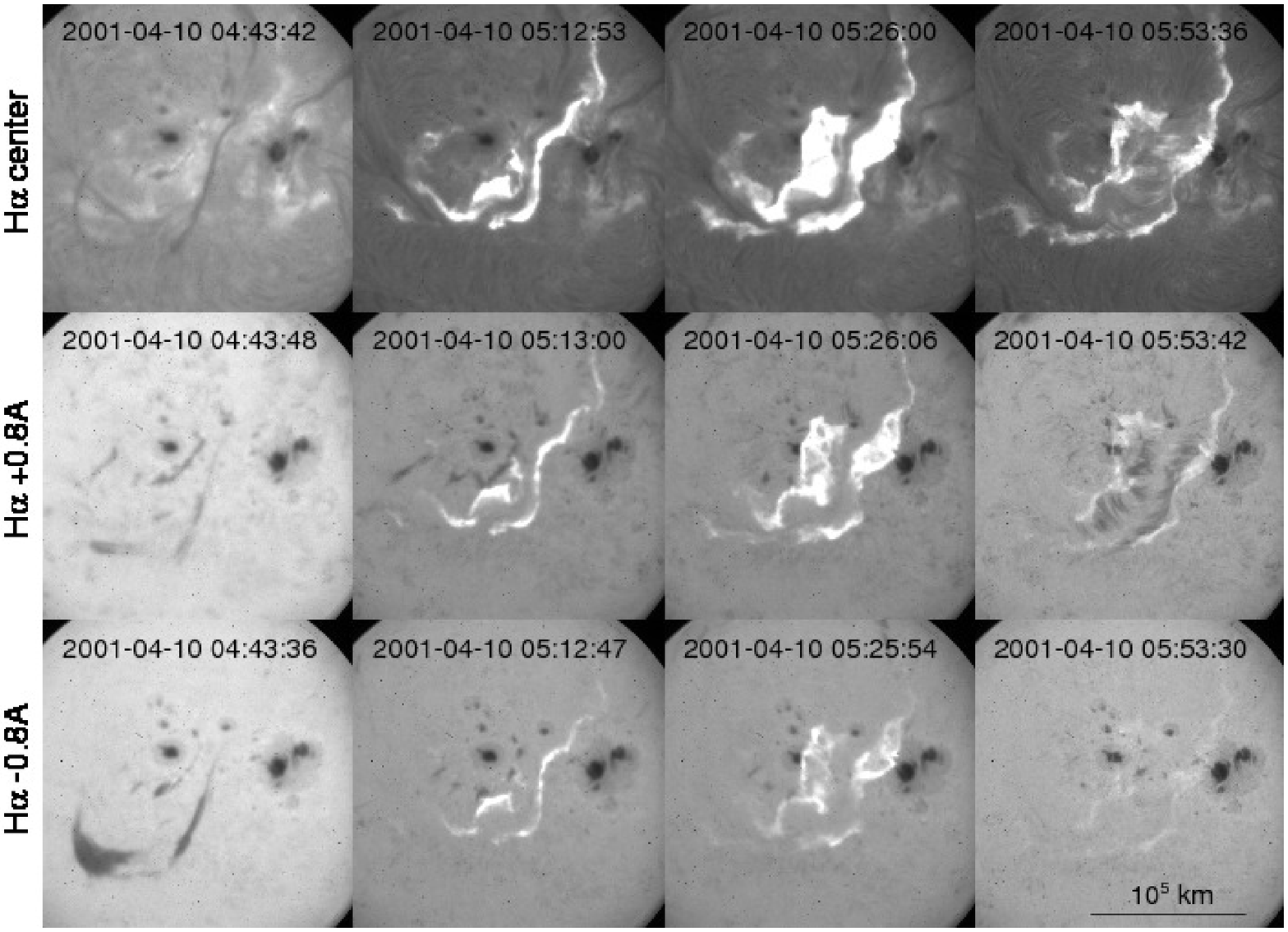}
  \end{center}
  \caption{
Temporal evolution of the 2001 April 10 flare in H$\alpha$ line center
({\it top}), in $+0.8$ {\AA} ({\it middle}), and in $-0.8$ {\AA} ({\it
bottom}).
Celestial north is up, and west is to the right.
}\label{fig:dst_ev}
\end{figure}

\subsection{Determination of Line Center}
Before examining the characteristics of the red asymmetry, we checked
the central position of the wavelength scan by the Lyot filter with
respect to the averaged H$\alpha$ line center of the target region.
Since not only the tuning offset of the filter but also the
solar rotation and the motion of the earth can cause a relative offset
of the wavelength, we should know how much the line center is shifted
during the observation.
We used the averaged intensity of a quiet region near the flare site
shown with the box in Figure~1 (a), and confirmed that the center of the
Lyot filter was $+0.05$~{\AA} displaced from the actual center of the
H$\alpha$ absorption.
Although such small offset may produces a fictitious shift of the
spectral line, it can hardly produce a significant asymmetry of
intensity as observed in far wings (such as at $\pm1.5$~{\AA}) of the
emission H$\alpha$ line.

Comparing Figure~1 (a) with (e), we can clearly see that the flare
ribbons are much brighter in the $+1.5$~{\AA} wing than in the
$-1.5$~{\AA} wings, and we can conclude that the intensity asymmetry is
mainly caused by the effect of the red asymmetry (or red shift) of the
H$\alpha$ line.
Thus, we neglect the small offset of the observing wavelength in the
following discussions.

\section{Red Asymmetry}
As the pair of the snap shots of Figure~1 (a) and (e) simply but clearly
show, most portions of the flare ribbons are brighter in the red-wing
(that is, H$\alpha$ $+1.5$~{\AA}) than those in the blue-wing (H$\alpha$
$-1.5$~{\AA}).
This means that red asymmetry appears all over the flare ribbons, not
only at bright kernels made by the bombardment of nonthermal particles,
but also at other fainter points that are likely generated by the
thermal conduction.

\subsection{Temporal Evolution of $RA$}
The right two panels of Figure~3 show the temporal evolution of the
H$\alpha$ intensities in the red ($+1.5$~{\AA}: red line) and blue
($-1.5$~{\AA}: blue line) wings for the selected flare kernels marked in
the left panel of Figure~3.
At these flare kernels, the intensity in the red wing $I_{red}$ is
always brighter than that in the blue wing $I_{blue}$ during the burst.
As we showed in the previous papers \citep{Asai03,Asai04}, we found
nonthermal emissions in hard X-rays and in microwaves observed with the
Hard X-ray Telescope (HXT; \cite{Kosu91}) on board {\it Yohkoh}
\citep{Oga91} and with the Nobeyama Radioheliograph (NoRH;
\cite{Naka94}), respectively, at these flare kernels.
We also plotted the light curves of the microwave (17~GHz) and the hard
X-ray (the H band; 53 -- 93~keV) emissions in Figure~3 with the black
and gray solid lines.
There are several peaks and the positions of the emission sources
move during the impulsive phase.
The flare kernels $\triangle$ and $\times$ are associated with the
nonthermal emission sources whose peak times are 05:19 and 05:22UT,
respectively, as colored with vertical gray regions in the right panels
of Figure~3.

Here, we note that both $I_{red}$ and $I_{blue}$ are subtracted with the
averaged intensity of the pre-brightening phase.
Both the intensities are set to be about zero, which is shown with the
horizontal dotted line in the right panels of Figure~3.
In the current case, the time gaps to take images at these two
wavelengths are also corrected.
The H$\alpha$ red ($+1.5$~{\AA}) wing images were always taken about
12~s previous to the paired blue ($-1.5$~{\AA}) wing images.
To make the time profiles of red asymmetry on the right panels of
Figure~3, we linearly interpolated the blue wing data so that the times
correspond the observing times of the red wing data.
We also have to note that the intensities are calculated by averaging
the region of 5$\times$5 pixels (which is equal to
$1.^{\prime\prime}4\times1.^{\prime\prime}4$) around the points marked
with the $\times$ and $\triangle$ signs.
These time profiles contain the region where the data suffer from the
saturation, and therefore, their peaks are possibly plateaued
(underestimated).
We showed the points which contain such saturated data with the upward
arrows in the right panels of Figure~3.

\begin{figure}
  \begin{center}
    \FigureFile(150mm,75mm){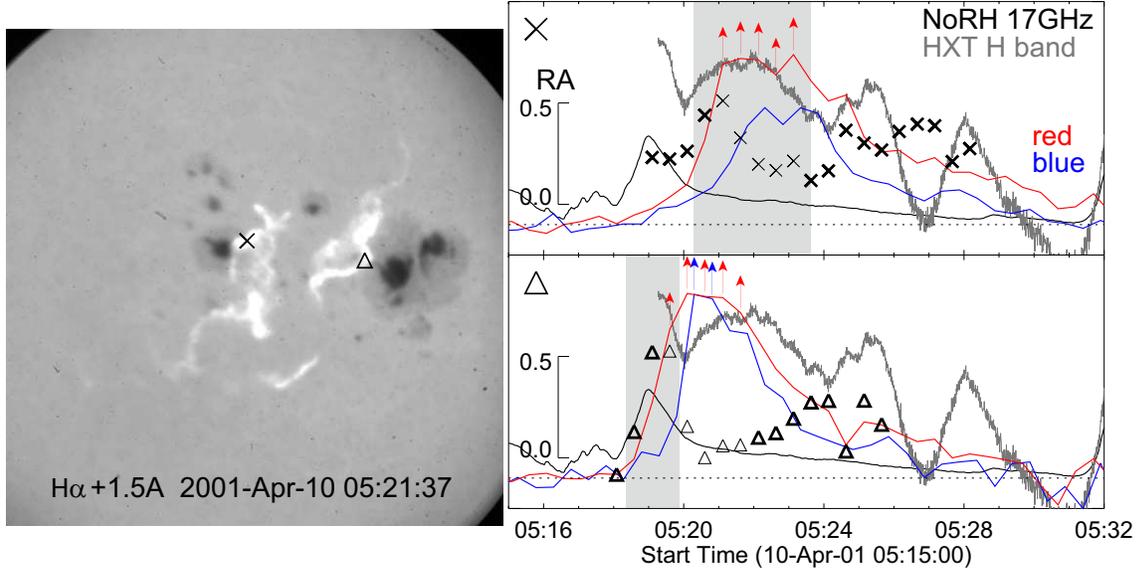}
  \end{center}
  \caption{
Temporal evolution of H$\alpha$ kernels of the flare.
The right panels show the time profiles of the H$\alpha$ kernels in the
red ($+1.5$~{\AA}; red line) and blue ($-1.5$~{\AA}; blue line) wings.
Microwave correlation plot observed at 17~GHz with NoRH and hard X-ray
count rate measured in the H-band (53 -- 93~keV) of {\it Yohkoh}/HXT
are also plotted with the black and gray solid lines, respectively.
The positions of the kernels are shown in the left panel with the cross
$\times$ and triangle $\triangle$.
Temporal evolution of $RA$ is also over-plotted with the same ($\times$
and $\triangle$) signs in the right panels.
The thin signs are the points calculated using the saturating
intensities.
Some points of the time profiles, especially the red-wing profiles,
suffer from the saturation, whose intensities could be underestimated as
shown with the upward arrows.
The horizontal dotted lines are the zero levels of the subtracted
intensities at the H$\alpha$ kernels.
}\label{fig:plot}
\end{figure}

From the intensity profiles, we also obtained the temporal evolution
of the red asymmetry ($RA$) by calculating 
\begin{equation}
RA = \frac{I_{red} - I_{blue}}{I_{red} + I_{blue}},
\end{equation}
where $I_{red}$ and $I_{blue}$ are intensities at the red- ($+1.5$~{\AA})
and blue- ($-1.5$~{\AA}) wings.
The temporal evolution of $RA$ is over-plotted with the $\times$ and
$\triangle$ signs in the right panels of Figure~3.
The data points suffering from the saturation are shown with the thin
sings in the plots.
These temporal evolutions of $RA$ imply that the peaks of the red
asymmetry precede those of the intensity by about 30 sec -- 1 minute,
while we cannot determine the exact peak times due to the saturation.
Moreover, the time profiles of $RA$ keep high values of about 0.3 -- 0.4
for several minutes even after the peaks.
By comparing the time profiles of $RA$ with the microwave and hard X-ray
emissions during the burst times, we found that the start times of the
$RA$ enhancements correspond with or even precede to those of the
nonthermal emissions.

\subsection{Spatial Distribution of $RA$}
We examined the spatial distribution of the red asymmetry $RA$, by using
the H$\alpha$ intensity in the red- and blue-wing images.
Both $I_{red}$ and $I_{blue}$ are subtracted with the averaged intensity
of the quiet region.
The quiet region used for the analysis is the same as that used above,
and is shown with the box in Figure~1 (a).
We present some snap shots of the 2-dimensional $RA$ maps to show in
Figure~4.
The red/blue color shows the region where red/blue asymmetry appears.
We can again confirm that the red asymmetry appears all over the flare
ribbons during the impulsive phase of the flare.
The green-hatched regions suffer from the saturation, and therefore, we
excluded them from the calculations of $RA$ here.
The temporal evolution of the $RA$ map is also shown.

\begin{figure}
  \begin{center}
    \FigureFile(140mm,140mm){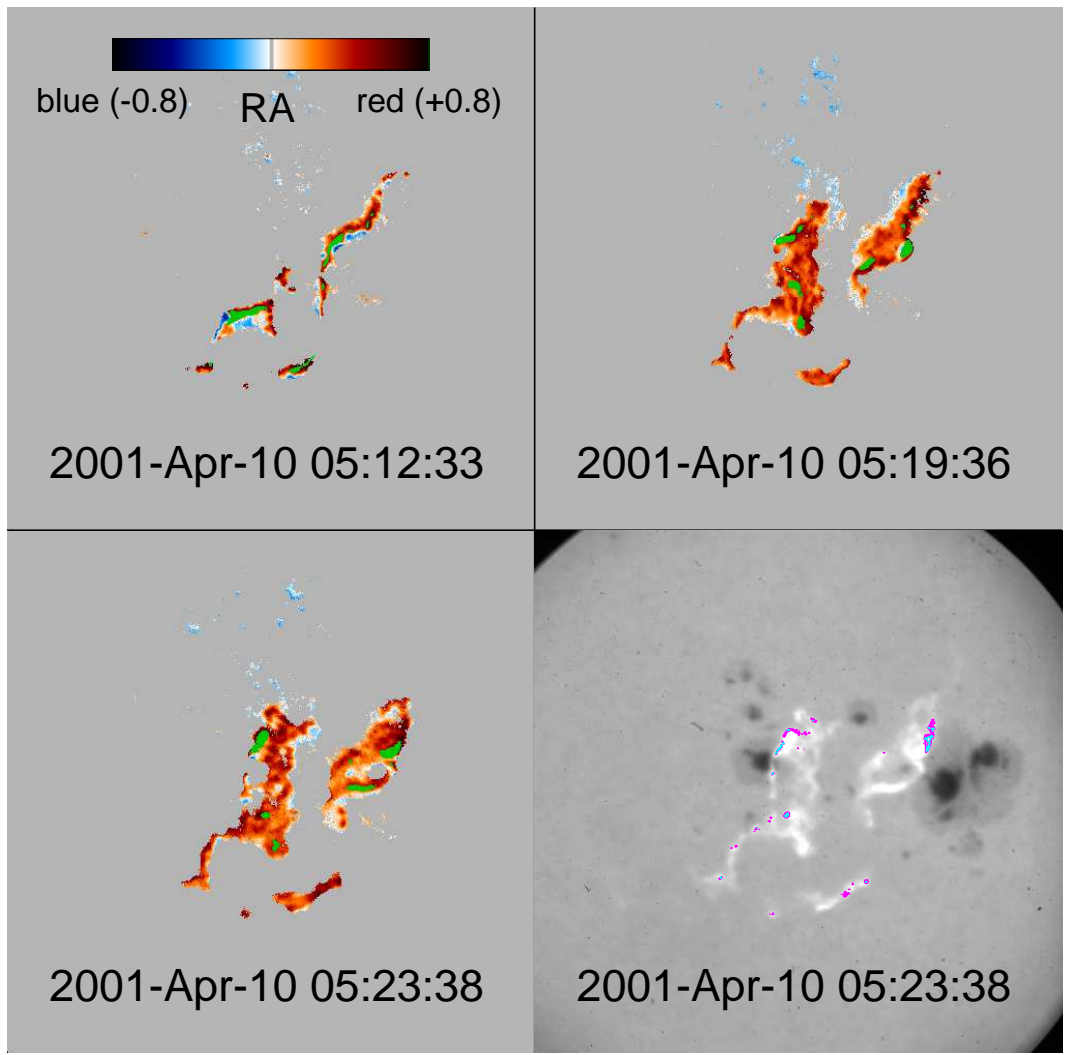}
  \end{center}
  \caption{
Evolution of red asymmetry $RA$ of the 2001 April 10 flare.
The regions where the H$\alpha$ data suffer from the saturation are
hatched with green color.
The bottom right panel shows an H$\alpha$ red-wing ($+1.5$~{\AA}) image
overlaid with the strong $RA$ ($>75 />85$~\%) positions with
magenta/light blue colors.
The times show those of the red wing data, and those for the blue wing
data are delayed about 12~s from them.
}\label{fig:asym}
\end{figure}

Here, we used only two wing data ($\pm$ 1.5{\AA}), and we did not follow
the whole line profiles, which could be strongly distorted during the
flare.
The $RA$ calculated in this paper, however, can be a representative
measure of the velocity, at least, at some layer of the chromosphere.
Since the emission at $\pm$ 1.5{\AA} is, more or less, optically thin
and is less affected from the radiative transfer effect of the
vertically stratified atmospheric condensation, it serves as a simple
measure of the line of sight velocity.
According to H$\alpha$ profiles of flares (e.g. \cite{Ichi84}), $RA$ at
around $\pm$ 1.5{\AA} certainly represent the Doppler shift in line
wings.

For the image taken at 05:23~UT, we also show the strong $RA$ position
($RA \gtrsim 0.57$ and 0.65; which corresponds to more than 75 and 85~\%
of the maximum value of $RA$) on the flare ribbon with the magenta and
light blue colors in the bottom right panel of Figure~4.
It shows that the strong $RA$ regions are preferentially located at the
outer edge of the flare ribbons.
The outer edges of flare ribbons are the footpoints of the ``newly
reconnected'' flare loops, in the context of the magnetic reconnection
model.
Therefore, our result means that strong $RA$ is associated with the
footpoints of such newly reconnected loops that probably provide strong
energy.
The concentration of strong $RA$ regions on the outer edges of flare
ribbons is consistent with the previous works (e.g. \cite{Fal97}).
The width of the strong $RA$ regions are about $1.^{\prime\prime}5$ --
$3.^{\prime\prime}0$ (1000 -- 2000~km).
The length along the direction of the flare ribbons can be longer, and
is about $1.^{\prime\prime}5$ -- $15^{\prime\prime}$ (1000 --
10000~km).
The largest ones are roughly consistent with the length of the hard
X-ray emission sources along the flare ribbons ($10^{\prime\prime}$ --
$20^{\prime\prime}$) that appeared during the impulsive phase of the
flare \citep{Asai04}.

As we already mentioned, there is time differences between the H$\alpha$
red wing data and the paired blue wing data.
It took about 12~s to shift the filter from the H$\alpha$ red
($+1.5$~{\AA}) wing to the blue ($-1.5$~{\AA}) wing (the red ones always
precede from the blue ones in the current case).
The flare ribbons expand during the flare with the speed of about 10 --
80 km s$^{-1}$, and even faster in the earlier phase (see \cite{Asai04},
for more detail).
Therefore, the positions of the outer edges of the flare ribbons
slightly different.
In the current case, the flare ribbons in the blue wing data are always
located outside (of about 100 -- 1000~km) of those seen in the red wing
data, which could cause ``fake'' blue asymmetry.
Indeed, we can recognize the blue asymmetry in the earlier phase as seen
in the top left panel of Figure~4.
Although \citet{Sve76} reported that such blue asymmetry sometimes
appear at the earlier phase of flares, in the current case, it may be
because the separation speed of the flare ribbons is much faster in the
earlier phase.
As the other panels of Figure~4 show, on the other hand, the time
difference is not important so much in the later phase.
Therefore, the RA maps in Figure~4 were made, ignoring the time
differences of the 12~s.
The times shown in Figure~4 are those for the red wing data.

\subsection{Scatter Plot}
We examined the relationship between the red asymmetry and the intensity
of the H$\alpha$ kernels.
Before this, we defined the summation ($SM$) and the difference ($DF$)
of the intensities at both wings as $SM = I_{red} + I_{blue}$ and $DF =
I_{red} - I_{blue}$.
Therefore, the red asymmetry $RA$ is consistent with $DF$ divided by
$SM$.
Again, we note that $I_{red}$ and $I_{blue}$ are subtracted with the
averaged intensity of the quiet region indicated with the rectangle in
Figure~1(a).

The left panel of Figure~5 shows the scatter plot between the difference
$DF$ (horizontal axis) and the summed H$\alpha$ intensity $SM$ (vertical
axis) at the kernels.
The ``kernels'' mentioned here are defined as the regions where the
intensity exceeds the average intensity at the quiet region, and
therefore, $SM$ exceeds 0.
We can clearly see the tendency that the stronger $DF$ is associated
with the brighter H$\alpha$ kernel.
The regions hatched with gray color are excluded from the discussion
because of the saturation.

The scatter plot shows that $SM$ is linearly proportional to $DF$, that
is, $SM = a \times DF~(+ b)$ with a proportional constant $a$ and an
intercept $b$.
By fitting the data, we get the constant $a$ to be about 2.5.
We also confirmed that the intercept $b$ is small (about 20 counts), and
we will ignore $b$ in the following discussion.
The dark gray line in the left panel of Figure~5 is the result of the
liner fitting to the data.

We also compared $RA$ with the H$\alpha$ intensity $SM$ as shown in the
right panel of Figure~5.
The horizontal axis is $RA$, while the vertical axis is the same as the
left panel.
The regions colored gray again show the regions where the kernels suffer
from the saturation.
The data points seems to scatter around a mean value of 0.37 without
significant correlation with intensity.
By using the constant $a$, we can derive the relation $(I_{red} -
I_{blue}) (I_{red} + I_{blue})^{-1} = DF/SM = RA = a^{-1}$, and
therefore, $RA = 0.4$ is expected.
The mean value of the line shift at different intensity does not vary
with the intensity.
The averaged value of about 0.37 is calculated for the bright kernels
highlighted with the black signs in the right panel of Figure~5 (that
is, the data points colored with right gray are excluded for the
calculation).

\begin{figure}
  \begin{center}
    \FigureFile(150mm,75mm){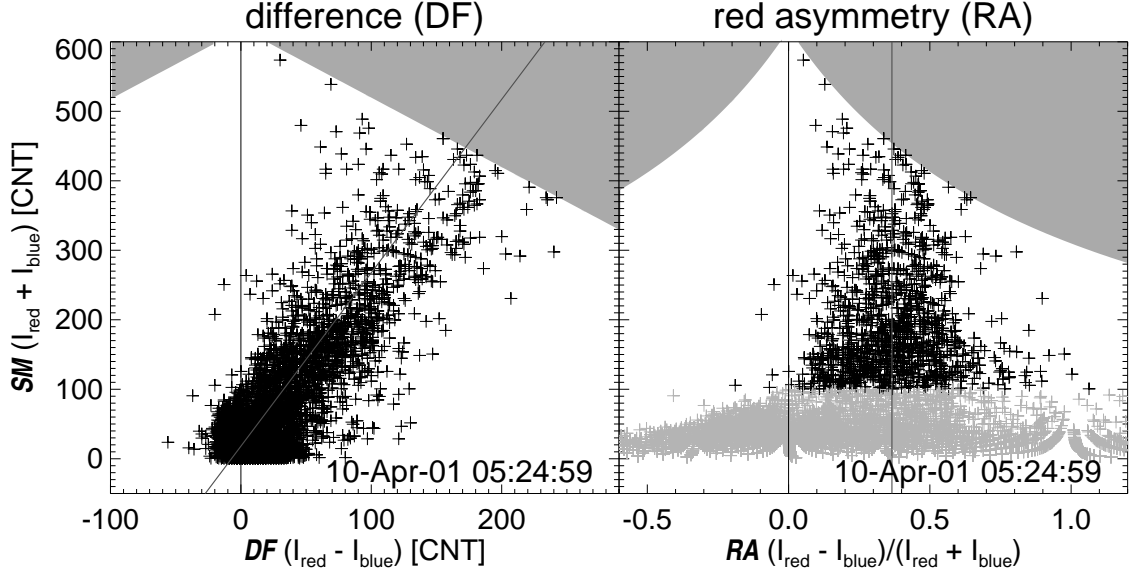}
  \end{center}
  \caption{
Scatter plot between the red asymmetries and the intensities of the
H$\alpha$ kernels.
Left: The horizontal axis shows the difference between $I_{red}$ and
$I_{blue}$ $DF$ ($= I_{red} - I_{blue}$).
The vertical axis shows their summed intensity $SM$ ($= I_{red} +
I_{blue}$).
The dark gray line is the result of the liner fitting of the data.
Right: The horizontal axis shows the red asymmetry $RA$, 
$(I_{red} - I_{blue})(I_{red} + I_{blue})^{-1} = DF/SM$.
The vertical axis is the same as that of the left panel.
The averaged value of the red asymmetry for the bright kernels (marked
with the black signs) is shown with the vertical gray line.
}\label{fig:scat}
\end{figure}

\section{Summary and Discussions}
We examined the spatially resolved red asymmetry distributions seen in
the 2001 April 10 solar flare and their temporal evolution, by using the
H$\alpha$ wing images taken with DST at Hida Observatory.
We found that red asymmetry appears all over the flare ribbons, both for
bright kernels generated by the bombardment of nonthermal particles and
for other fainter regions generated by the (thermal) conduction.
We confirmed that the strong $RA$ appears at the outer edge of the flare
ribbons that are the footpoints of newly reconnected flare loops.
The width of about 1.5 -- 3.0$^{\prime\prime}$ ($\sim$ 1000 -- 2000~km).
This supports that the strong energy release occurs on the magnetic
loops that connect the outermost edges of the flare ribbons.
We found a clear tendency that the larger difference between the
H$\alpha$ red and the blue wing intensities $DF$ is associated with the
brighter H$\alpha$ kernels $SM$.
It turns out to be that the red asymmetry ($RA = (I_{red} - I_{blue})
(I_{red} + I_{blue})^{-1} =DF/SM$) does not depend on the intensity of
the H$\alpha$ kernels, and scatters around a mean value of about 0.37.

In Figure~3 we showed that the temporal evolution of $RA$ does not
synchronize with that of the intensity, and the peak times of $RA$
time profiles precedes those of the intensities.
Moreover, the enhancements of $RA$ time profiles procede to those of the
nonthermal emissions observed in hard X-rays and/or in microwaves.
\citet{Zarro89} also mentioned that redshift in the H$\alpha$ wings
starts to appear before the enhancement of the hard X-ray emission, that
is, strong energy release occurs.
At flare kernels, not only nonthermal particles but thermal conduction
precipitate from the energy release site in the corona into the
chromosphere.
The preceding enhancement of $RA$ to the hard X-ray emission could be
caused by such other mechanisms rather than by nonthermal particles.

To make clear the temporal evolutions of $RA$ and intensity, we overlaid
the temporal trajectories of the H$\alpha$ kernels, shown in Figure~3
with the $\times$ and $\triangle$ sings, on the scatter plot between
$RA$ and the summed intensity $SM$, which is presented as Figure~6.
The $RA$ plots move counter-clockwise on the scatter plot, and after the
$RA$ enhancements, the intensities reaches the peak.
The absence of a clear correlation with the intensity is because the
time profile of $RA$ for individual kernels evolves independently with a
manner that the peak downward velocity is not a strong function of their
peak intensity.
\citet{Can90b} also reported that there is no one-to-one relation
between brightness of H$\alpha$ line center and redshift, while they
found many redshift features at bright kernels at H$\alpha$ line center.

If we assume a constant width of the H$\alpha$ emission line, we can
regard $RA$ as an indicator of the velocity.
For example, assuming a gaussian function with the 
full-width-half-maximum $FWHM$ in {\AA}, $RA$ can be related to the
Doppler (redward) velocity as $2.75 \times \ln{(1+RA)(1-RA)^{-1}} (FWHM
({\rm{\AA}}))^2$~km~s$^{-1}$.
The estimation of the width of the H$\alpha$ emission line is, however,
very difficult, because the shape of the line is drastically changed (it
cannot be fitted with a gaussian function) and especially the Stark
effect broadens the line at the wings \citep{Sve76}.
From the results of \citet{Ichi84} (in the Figs. 3b and 4b), we
estimated $RA$ for their flares, and they are about 0.2 -- 0.5, which is 
comparable to the 2001 April 10 flare.
They also reported that the Doppler velocities are 10 --
100~km~s$^{-1}$, the FWHM of about 4.5 -- 6.0 is expected.
If we assume that the FWHM is about 5.0, here, $RA$ of 0.37 turns out
the Doppler (downward) velocity of about 53~km~s$^{-1}$.
Scattering of $RA$ around a certain value of about 0.37 is also because
the time profiles of $RA$ keep high values for several minutes even
after the intensity peaks.
This means that the downward motion at flare ribbons continues with the
velocity of about several tens of km~s$^{-1}$ independently from the
intensity.

\begin{figure}
  \begin{center}
    \FigureFile(150mm,75mm){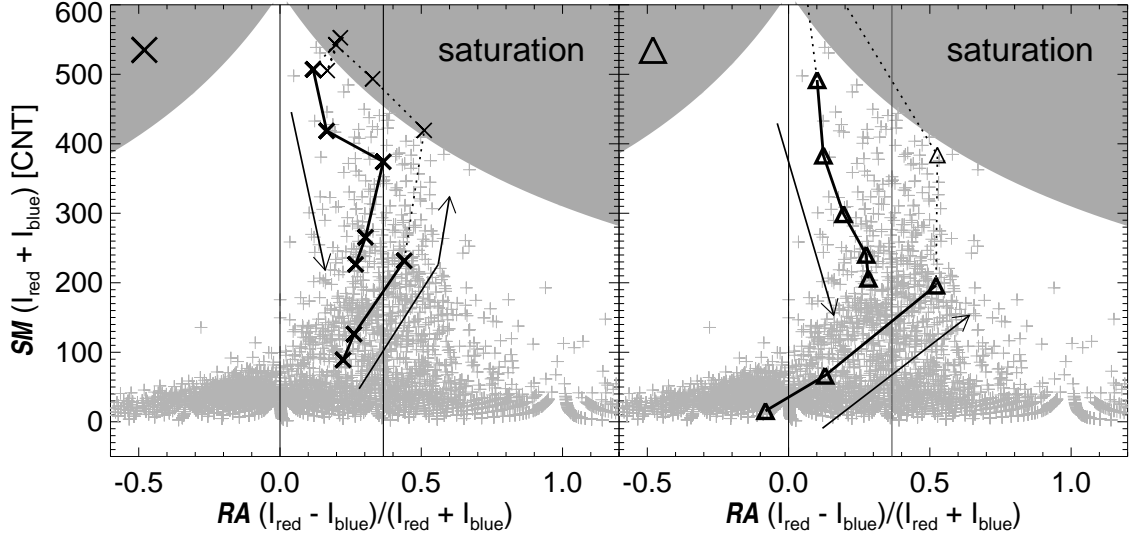}
  \end{center}
  \caption{
Temporal trajectories of H$\alpha$ kernels in the $RA$ - $SM$ scatter
plot.
The temporal evolutions of $RA$ are shown in the right panels of
Figure~3, and the marks of the H$\alpha$ kernels (cross $\times$ and
triangle $\triangle$ signs) are also the same as those shown in
Figure~3.
The arrows guides the temporal evolutions.
The background $RA$ -- $SM$ scatter plots are the same as the right
panel of Figure~5.
}\label{fig:scat}
\end{figure}

\bigskip 

Acknowledgements

We first acknowledge an anonymous referee for her/his useful comments
and criticisms.
This work was supported by the Grant-in-Aid for the Global COE Program
``The Next Generation of Physics, Spun from Universality and Emergence''
from the Ministry of Education, Culture, Sports, Science and Technology
(MEXT) of Japan.

{}

\end{document}